\documentclass[man]{apa6}

\usepackage[american]{babel}
\usepackage{csquotes}
\usepackage[style=apa,sortcites=true,sorting=nyt]{biblatex}
\usepackage{graphicx,epsfig}
\usepackage{amsmath}
\usepackage{amssymb}
\usepackage{color}
\usepackage[makeroom]{cancel}
\usepackage{verbatim}
\usepackage{bm}
\usepackage{fancybox}
\usepackage{color}
\usepackage{mdframed}
\usepackage{tikz}
\usepackage{caption}
\usepackage{subcaption}
\usepackage{threeparttable}
\usepackage{listings} 
\usetikzlibrary{shapes.geometric, arrows} 
\usepackage{multirow}
\usepackage{threeparttable}
\usepackage{hyperref}
\usepackage{hyperref}
\usepackage{hyperref}
\hypersetup{
    colorlinks=true,    
    linkcolor=black,    
    citecolor=black,    
    filecolor=black,    
    urlcolor=black      
}

\DeclareLanguageMapping{american}{american-apa}
\title{A Comparison of Full Information Maximum Likelihood and Machine Learning Missing Data Analytical Methods in Growth Curve Modeling}
\shorttitle{Missing Data Analytical Methods}
\author{Dandan Tang and Xin Tong}
\affiliation{Department of Psychology, University of Virginia}
\authornote{This research was supported by a 4-VA grant from the Commonwealth of Virginia. Correspondence concerning this article should be addressed to Dandan Tang, Department of Psychology, University of Virginia. Email: gux8df@virginia.edu.}
\abstract{Missing data are inevitable in longitudinal studies. Traditional methods, such as the full information maximum likelihood (FIML), are commonly used to handle ignorable missing data. However, they may lead to biased model estimation due to missing not at random data that often appear in longitudinal studies. Recently, machine learning methods, such as random forests (RF) and K-nearest neighbors (KNN) imputation methods, have been proposed to cope with missing values. Although machine learning imputation methods have been gaining popularity, few studies have investigated the tenability and utility of these methods in longitudinal research. Through Monte Carlo simulations, this study evaluates and compares the performance of traditional and machine learning approaches (FIML, RF, and KNN) in growth curve modeling. The effects of sample size, missingness rate, and missing data mechanism on model estimation are investigated. Results indicate that FIML is a better choice than the two machine learning imputation methods regarding model estimation accuracy and efficiency.} 
\begin{document} 
\maketitle

\section{Introduction}
In the realm of longitudinal research, the challenge of missing data is ubiquitous and can greatly influence the accuracy and reliability of research outcomes. To address problems related to missing data, many statistical methods have been developed based on the underlying mechanisms that may cause the missingness, termed missing data mechanisms (Little \& Rubin, 2002). The three categories of the missing data mechanisms are missing completely at random (MCAR), where the data's absence is unrelated to both observed and unobserved values; missing at random (MAR), where the missing data is contingent on the observed values but not on the missing values themselves; and missing not at random (MNAR), where the absence is directly tied to the missing values even when considering the observed data. Since MCAR data can be effectively handled by all existing missing data analytical methods, this paper focuses on statistical methods for analyzing MAR and MNAR data in longitudinal research.  

Many traditional statistical methods exist for the analysis of missing data, among which the full information maximum likelihood (FIML) stands out as one of the most popular tools (Enders, 2010; Rodriguez, 2023). FIML has been implemented in nearly every SEM program (Muthén \& Muthén, 2017;  Rosseel, 2012) because of two favorable properties: its ability to produce consistent parameter estimates and the efficiency of those estimates. However, these two properties may not hold under the MNAR cases (Yuan, 2009). 

Recently, machine learning techniques, including random forests (RF) and K-nearest neighbors (KNN) imputation methods, have been developed as powerful tools to cope with missing values (Hayes et al., 2015; Stekhoven \& Bühlmann, 2012). Unlike traditional methods, these approaches operate without specific distributional assumptions regarding the data and thus the exploration of these methods within the context of longitudinal research remains limited, with very few studies delving into the tenability and utility of these techniques under the MAR and MNAR missing data mechanisms (Hayes \& McArdle, 2017).   

Despite the prevalence of missing data in longitudinal research and machine learning techniques, the gap in understanding machine learning analytical methods for handling missing data in longitudinal studies signifies an important area for in-depth examination. Thus, this research aims to investigate the performance of traditional and machine learning analytical methods for longitudinal data with different missing data mechanisms. By methodically and numerically assessing the performance of these methods, the research intends to find both their practical applicability and potential limitations within longitudinal research contexts. 

\section{Growth Curve Models}
Growth Curve models (GCMs) are one of the most essential and frequently used models for fitting longitudinal data for which the same subjects are observed repeatedly over time. Suppose a cohort of $N$ individuals participates in a longitudinal study and is measured $T$ times repeatedly. Let $\textbf{y}_i = (y_{i1}, ..., y_{iT} )$ represent the observed scores for the $i$th individual, and $y_{it}$ be the observed value for the $i$th individual at measurement occasion t ($t= 1, ..., T$). An unconditional GCM can be typically expressed as
\begin{equation}
 \begin{aligned}
 & \textbf{y}_i = \mathbf{\Lambda}_i\textbf{b}_i + \textbf{e}_{i},	\\
 & \textbf{b}_i = \mathbf{\beta} + \textbf{u}_i,
 \end{aligned}
\end{equation}
where $\mathbf{\Lambda}_i$ is a $T$ x $q$ factor loading matrix recording the shapes of growth trajectories; the vector $\textbf{b}_i$ is a $q$ x $1$ vector of random effects, and $\textbf{e}_{i}$ is a vector of intraindividual measurement errors. The random effects $\textbf{b}_i$ vary across individuals, while its mean $\mathbf{\beta}$ represents the fixed effects for the population. The residual vector $u_i$ represents the random component of $\textbf{b}_i$. Generally, it is assumed that both $\textbf{e}_{i}$ and $\textbf{u}_i$ follow multivariate normal (MN) distributions, such that
\begin{equation}
 \begin{aligned}
 & \textbf{e}_{i} \sim \mathcal{MN}_T(0,\mathbf{\Phi}),\\
 & \textbf{u}_i \sim \mathcal{MN}_q(0,\mathbf{\Sigma}),
 \end{aligned}
\end{equation}
where the subscript of the MN distribution implies the dimensionality of the random vector. $\Phi$ is a $T$ x $T$ covariance matrix of $e_i$ and is usually assumed to be diagonal $\Phi = \sigma^2_eI$. This assumption indicates that intraindividual measurement errors have equal variance and are independent across different measurement occasions.

\section{Missing Data Analytical Methods}
When growth curve modeling is used to handle incomplete longitudinal data, missing data analytical methods should be applied. This section briefly introduces full information maximum likelihood, random forests, and K-nearest neighbors imputation methods, whose performance will be evaluated and compared in the Monte Carlo simulation study.   

\subsection{Full Information Maximum Likelihood Method}
The full information maximum likelihood method fully uses the available data, including partially missing or fully observed, to produce parameter estimates that maximize the likelihood function (Enders \& Bandalos, 2001).  Given a set of parameters, the likelihood function essentially represents the "fit" of a statistical model to the observed data. Studies showed that FIML can provide unbiased estimates for normally distributed data under the MAR mechanism (Enders \& Bandalos, 2001). 

\subsection{Random Forest Imputation Method}
The presence of missing data reduces statistical power and leads to potential bias. Therefore, many imputation techniques have been developed over the years to deal with missing data. Random forest imputation is a machine learning based method that was developed as a robust, nonparametric alternative to traditional methods. This imputation method deals with missing data by predicting missing values based on other observed variables using a random forest trained on the complete cases in the dataset (Stekhoven \& Bühlmann, 2012). Once the missing values of one variable have been estimated and filled in, this "new" data can be used as part of the predictors for estimating the missing values of other variables (Doove, Van Buuren, \& Dusseldorp, 2014). This process is repeated as the new estimates contribute to the accuracy of subsequent estimates. When certain stopping criteria are reached (e.g., maximum number of iterations or convergence of the estimates to a stable solution), the process stops.

This RF imputation method is popular because of its unique advantages. First, it does not rely on any distributional assumptions about the data. Moreover, it can handle different types of data - numerical, categorical, or a mixture of both - without any preliminary transformations. Furthermore, this method can handle high-dimensional datasets, capturing complex interactions and nonlinear relationships in the data (Stekhoven \& Bühlmann, 2012). Such advantages have promoted the application of the RF imputation, as practical data nowadays are often collected from various sources and can be quite complex. However, we would like to note that the RF imputation can be computationally intensive, especially for large datasets. 

\subsection{K-nearest Neighbors Imputation Method}
The K-nearest neighbors imputation is also a machine learning based method. It handles missing data based on the observed data that are most similar to them. The process of this method mainly includes the following two steps. First, "k" observations (called k-nearest neighbors) that are most similar to the missing points are identified based on some features (Jonsson \& Wohlin 2004). For example, Euclidean distance can be used to determine observations' nearness. Second, these "k" observations are employed to generate an imputed value for the missing point through a simple mean or weighted average (Batista \& Monard, 2002). 

The KNN imputation also does not require any distributional assumptions about the data and can simultaneously handle categorical and numerical variables. Moreover, because KNN relies on local information around missing points, it often generates more accurate estimates than methods using global information, like mean or median imputation (Batista \& Monard, 2002). However, this method may be very sensitive to the choice of k. If k is too large, neighbors that are not truly similar or relevant might influence estimation, leading to less accurate estimates (Zhang et al., 2017). If k is too small, the imputation is highly sensitive to minor changes in neighbors, leading to unstable estimates.

\section{A Simulation to Compare the Performance of FIML, RF, and KNN}
\subsection{Simulation Design}
The aim of this section is to evaluate the performance of FIML, random forests imputation, and KNN imputation for longitudinal missing data analysis via a Monte Carlo simulation study. The data were generated from a linear growth curve model with four measurement occasions. Following Tong, Zhang, and Yuan (2014),  the population parameters in the model were set as below. The fixed effects of the latent intercept and slope were 6 and 2 ($\beta=(\beta_L, \beta_S)'= (6, 2)'$), respectively. The variance of the latent intercept was 1 ($\sigma_L^2$), the variance of the latent slope was 1 ($\sigma_S^2$), and the correlation between the latent intercept and slope was 0. The variance of the intraindividual measurement errors was also set at 1 ($\sigma_e^2$ = 1).
 
In this simulation, we manipulated sample size (N = 100, 200, 300), rate of missingness (0\%, 5\%, 15\%, 30\%), and missing data mechanism (MAR, MNAR). For the MAR data,  the missingness in the outcome variables depends on the observations from the previous time points. For MNAR data, if an auxiliary variable \textit{A}, related to the latent slope, is larger than the given percentiles, the outcome variables are missing. All data were generated using R, and the code is provided on our GitHub site \href{https://github.com/DandanTang0/Missing-data-FIML-RF-and-KNN/tree/main/Data%20generation}{(https://github.com/DandanTang0/Missing-data-FIML-RF-and-KNN/tree/main/Data\%20generation)}.

For each data condition, a total of 500 datasets were generated. Each dataset was analyzed using the three missing data analytical methods. Note that the complete data were analyzed with the maximum likelihood estimation.  The R package \textit{lavaan} was used for missing data analysis with FIML (Rosseel, 2012). The random forests and KNN imputation methods were conducted using the \textit{missForest} and \textit{VIM} packages in R, respectively (Stekhoven \& Bühlmann, 2012; Templ et al., 2022).  

\subsection{Results}
To assess the performance of the three methods, we computed the relative bias to measure how much an estimate deviated from the true population value. We also computed the coverage rate of the parameter estimate confidence interval to measure the probability that the true value lay within the 95\% confidence interval among 500 replicates. Figure 1 presents the relative bias results for the estimated average of latent slopes ($\beta_S$), and Figure 2 presents the relative bias results for the estimated variance of the latent slopes ($\sigma_S^2$). Relative bias results for other model parameters (the average of the latent intercepts, the variance of the latent intercepts, and the correlation between the latent intercepts and slopes) and the confidence interval coverage rates results can be found on our GitHub site  \href{https://github.com/DandanTang0/Missing-data-FIML-RF-and-KNN/tree/main/simulationResults}{https://github.com/DandanTang0/Missing-data-FIML-RF-and-KNN/tree/main/simulationResults}.

Under the MAR mechanism, Figures 1 and 2 show that FIML outperforms both RF and KNN imputations in estimating parameters related to the latent slopes. The performance of FIML is also consistent across different sample sizes and missingness rate conditions. The results for other model parameters (in the supplemental results on our GitHub site) are consistent with those from the latent slopes. This consistency strongly suggests that FIML is the most effective method among the three for handling data with the MAR mechanism. 

Under the MNAR mechanism, the figures imply that FIML beats the other two methods in estimating the average of the latent slopes, while the performance of FIML in estimating the variance of the latent slopes is comparable to RF and KNN imputations. Also, the performance of the three methods deteriorates with an increasing missingness rate but slightly improves with a larger sample size. The results for other parameters (available on our GitHub site) are consistent with those from the latent slopes, where FIML outperforms or is comparable to the other two methods. Thus, in general, FIML stands out as the most suitable method among the three for handling MNAR missing data.

\begin{figure}[h]
    \centering
    \includegraphics[scale=0.6]{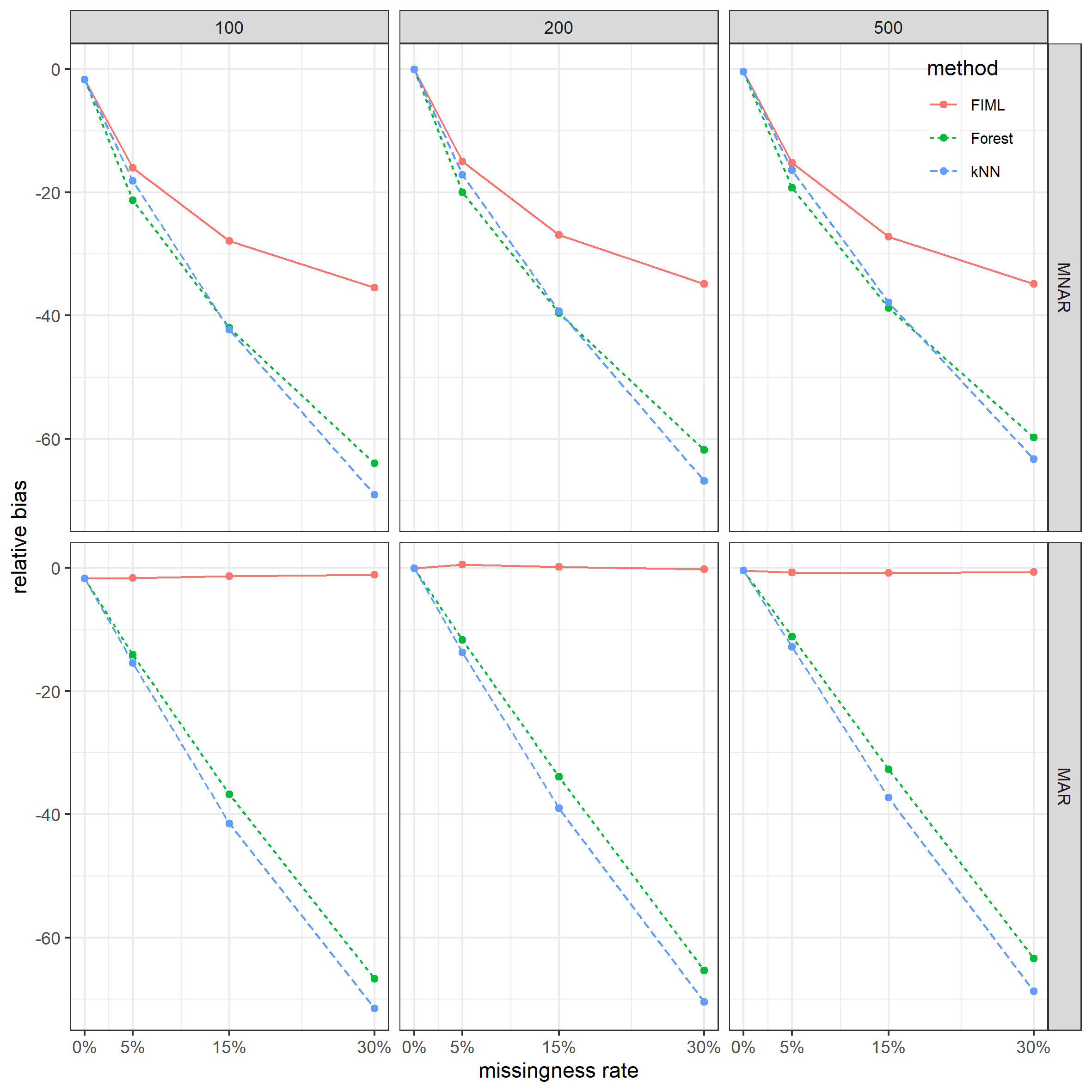} 
    \caption{The relative bias for the average of the latent slope estimates  ($\beta_S$)} 
    \label{fig:image1}
    
    \vspace{0.5cm} 
    
    \includegraphics[scale=0.6]{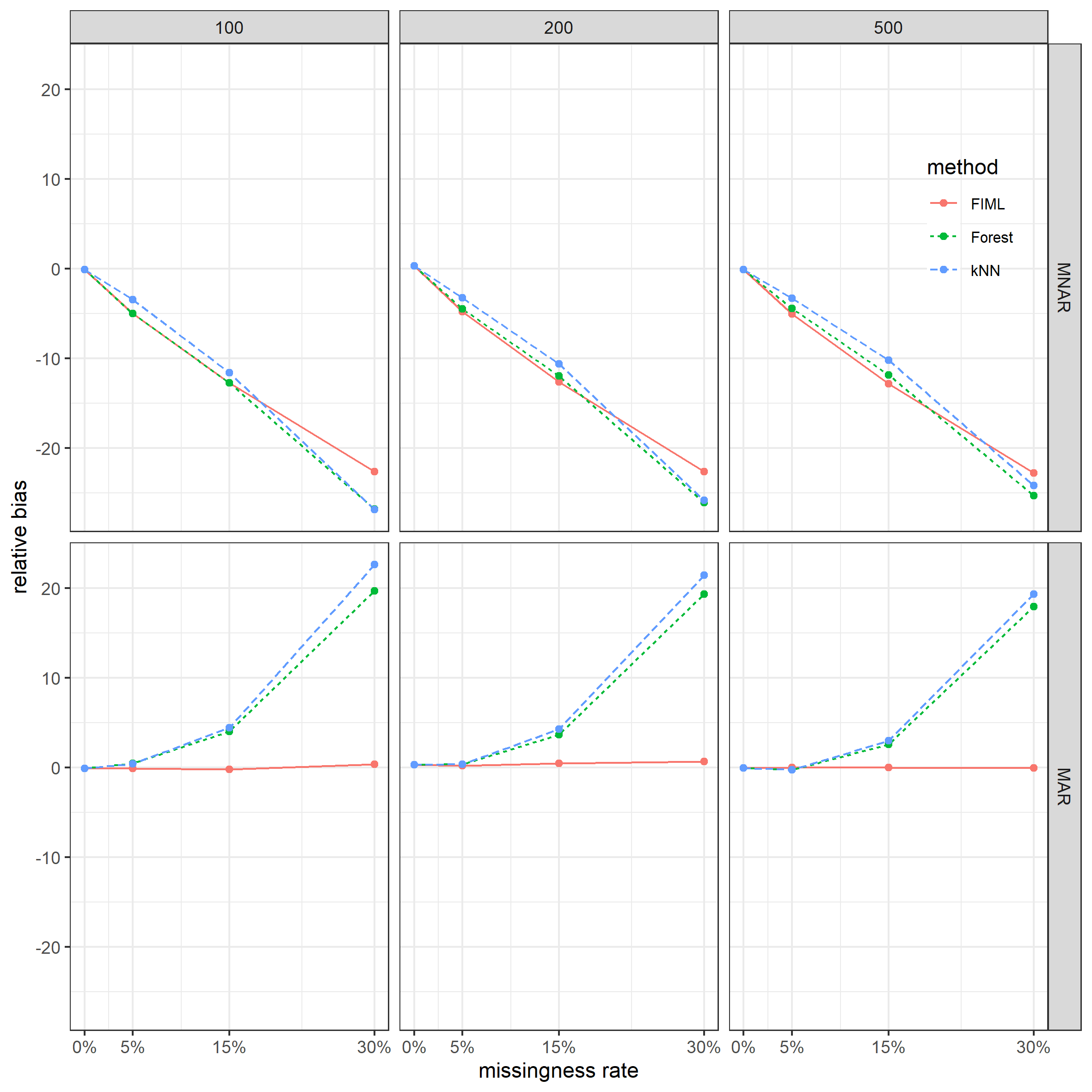} 
    \caption{The relative bias for the average of the latent slope's variance estimates ($\sigma_S^2$)} 
    \label{fig:image2}
\end{figure}

\section{Discussion and Conclusion}
Through a Monte Carlo simulation, this study evaluated FIML and two machine learning methods for handling missing data with MAR and MNAR missing data mechanisms in growth curve modeling. Contrary to prevailing perceptions regarding the efficacy of machine learning techniques, our findings indicated that FIML is the optimal one among the three.

Machine learning techniques have gained significant attention over the past few decades, often outperforming traditional methods across various domains. Yet, they appear to fail when addressing missing data in longitudinal studies. This deserves deeper exploration. One key distinction between traditional and machine learning methods lies in the assumption about data distribution. While traditional methods often rely on normal distributional assumptions, machine learning techniques generally do not have such assumptions. Given that our simulation focused solely on normal data, it is imperative to investigate the performance of machine learning and traditional methods in the context of non-normally distributed data. 

The machine learning methods used in this study are classified as single-imputation methods, wherein each missing value is filled with one estimated value. However, single imputation neglects the uncertainty associated with missing data, which multiple imputation methods aim to address (Van Buuren, 2018). Multiple imputations often yield more accurate and unbiased parameter estimates by generating more than one possible value for each missing data point. Future studies could, therefore, extend the scope of the current research by exploring the effectiveness of machine learning methods based on multiple imputations in longitudinal missing data analysis.

\section{References}
\begin{description}
\item Batista, G. E., \& Monard, M. C. (2002). A study of K-nearest neighbour as an imputation method. \emph{His}, 87(251-260), 48.
\item Doove, L. L., Van Buuren, S., \& Dusseldorp, E. (2014). Recursive partitioning for missing data imputation in the presence of interaction effects. \emph{Computational statistics \& data analysis, 72}, 92-104.
\item Enders, C. K. (2022). \emph{Applied missing data analysis}. New York, NY: Guilford.
\item Enders, C. K., \& Bandalos, D. L. (2001). The relative performance of full information maximum likelihood estimation for missing data in structural equation models. \emph{Structural equation modeling, 8}(3), 430-457.
\item Hayes, T., \& McArdle, J. J. (2017). Evaluating the performance of CART-based missing data methods under a missing not at random mechanism. \emph{Multivariate Behavioral Research, 52}(1), 113-114.
\item Hayes, T., Usami, S., Jacobucci, R., \& McArdle, J. J. (2015). Using Classification and Regression Trees (CART) and random forests to analyze attrition: Results from two simulations. \emph{Psychology and aging, 30}(4), 911.
\item Jonsson, P., \& Wohlin, C. (2004, September). An evaluation of k-nearest neighbour imputation using likert data. \emph{In 10th International Symposium on Software Metrics, 2004}. Proceedings. (pp. 108-118). IEEE.
\item Lei, P. W., \& Shiverdecker, L. K. (2020). Performance of estimators for confirmatory factor analysis of ordinal variables with missing data. \emph{Structural Equation Modeling: A Multidisciplinary Journal, 27}(4), 584-601.
\item Little, R. J. A., \& Rubin, D. B. (2019). \emph{Statistical analysis with missing data} (2nd ed.). New York, NY: Wiley-Interscience
\item Muthén, L. K., \& Muthén, B. (2017). \emph{Mplus user's guide: Statistical analysis with latent variables}
\item Rodriguez, D. (2023). Area under the Curve as an Alternative to Latent Growth Curve Modeling When Assessing the Effects of Predictor Variables on Repeated Measures of a Continuous Dependent Variable. \emph{Stats, 6}(2), 674-688.
\item Rosseel, Y. (2012). \emph{lavaan: a brief user’s guide}. 
\item Stekhoven, D. J., \& Bühlmann, P. (2012). MissForest—non-parametric missing value imputation for mixed-type data. \emph{Bioinformatics, 28}(1), 112-118.
\item Templ, M., Alfons, A., Kowarik, A., Prantner, B., \& Templ, M. M. (2022). Package ‘VIM’.
\item Tong, X., Zhang, Z., \& Yuan, K. H. (2014). Evaluation of test statistics for robust structural equation modeling with nonnormal missing data. \emph{Structural Equation Modeling: A Multidisciplinary Journal, 21}(4), 553-565.
\item Tong, X., Zhang, T., \& Zhou, J. (2021). Robust Bayesian growth curve modelling using conditional medians. \emph{British Journal of Mathematical and Statistical Psychology, 74}(2), 286-312.
\item Van Buuren, S. (2018). \emph{Flexible imputation of missing data}. 2nd edition, CRC press.
\item Yuan, K. H. (2009). Normal distribution based pseudo ML for missing data: With applications to mean and covariance structure analysis. \emph{Journal of Multivariate Analysis, 100}(9), 1900-1918.
\item Yuan, K. H., \& Zhang, Z. (2012). Robust structural equation modeling with missing data and auxiliary variables. \emph{Psychometrika, 77(4)}, 803-826.
\item Zhang, S., Li, X., Zong, M., Zhu, X., \& Wang, R. (2017). Efficient kNN classification with different numbers of nearest neighbors. \emph{IEEE transactions on neural networks and learning systems, 29}(5), 1774-1785.
\end{description} 

\end{document}